\documentclass[aps,prb,twocolumn,showpacs,preprintnumbers,groupedaddress]{revtex4}
\usepackage{graphicx} 

\begin{document}

\draft


\title{
Quantum wells with atomically smooth interfaces
}

\author{
Masahiro Yoshita${}^1$, Hidefumi Akiyama${}^{1,2}$, Loren N. Pfeiffer${}^2$, and Ken W. West${}^2$
}

\affiliation{
${}^1$Institute for Solid State Physics, University of Tokyo, \\
5-1-5 Kashiwanoha, Kashiwa, Chiba 277-8581, Japan\\ 
${}^2$Bell Laboratories, Lucent Technologies, 
600 Mountain Avenue, Murray Hill, NJ 07974, USA, \\
}

\date{\today}

\begin{abstract}
By a cleaved-edge overgrowth method with molecular beam epitaxy and a (110) growth-interrupt-anneal, we have fabricated a GaAs quantum well exactly 30 monolayers thick bounded by atomically smooth AlGaAs hetero-interfaces without atomic roughness. Micro-photoluminescence imaging of this quantum well indeed shows spatially uniform and spectrally sharp emission over areas of several tens of $\mu$m in extent.
By adding a fractional GaAs monolayer to our quantum well we are able to study the details of the atomic step-edge kinetics responsible for flat interface formation.
\end{abstract}

\maketitle 

\narrowtext

Molecular beam epitaxy (MBE) is well established as one of the premier techniques in modern nanostructure fabrication. However, on an atomic scale, all quantum wells (QWs) grown by conventional MBE have rough barrier-well interfaces with step-edges, pits, or islands one or more monolayers (MLs) high.\cite{HessSCI,ZrennerPRL,BrunnerPRL,GammonSCI,WuPRL} This hetero-interface roughness causes localization of the electronic states of electrons and holes in confined nanostructures such as quantum wells and wires into zero-dimensional (0-D) quantum dots. These localized quantum dots are useful in research on the fundamental nature of zero dimensions.\cite{GammonSCI} However, in quantum wells or wires the inherent 2-D or 1-D properties are disturbed by this atomic roughness and dynamics of the free 2-D or 1-D carriers or excitons are suppressed, preventing a full systematic understanding of semiconductor physics as a function of dimensionality.

Cleaved-edge overgrowth\cite{PfeifferAPL} is a novel method of GaAs MBE on the atomically flat (110) surface exposed by an $\it{in}$ $\it{situ}$ cleave so that there exists no roughness on this bottom interface. The top surface of the overgrown film, however, generally has large roughness due to the required GaAs growth conditions on the cleaved (110) surface.\cite{HasenNAT,YoshitaPRB} 

In this letter, we demonstrate complete removal of the surface roughness on a (110) GaAs epitaxial layer by growing on the cleave an integral number of GaAs MLs with an $\it{in}$ $\it{situ}$ growth-interrupt-anneal, and fabrication of a 6-nm (110) GaAs QW exactly 30-MLs thick without barrier-well interface roughness. Micro-photoluminescence (micro-PL) imaging and spectroscopy on this QW indeed shows a spatially uniform and spectrally sharp emission.
Preliminary experiments on (110) GaAs annealing were done using homoepitaxial overgrowth\cite{YoshitaJJAP}, however, this is the first hetero-epitaxial study and is the first fabrication of an atomically smooth QW without interface roughness.

\begin{figure}
\includegraphics[width=.30\textwidth]{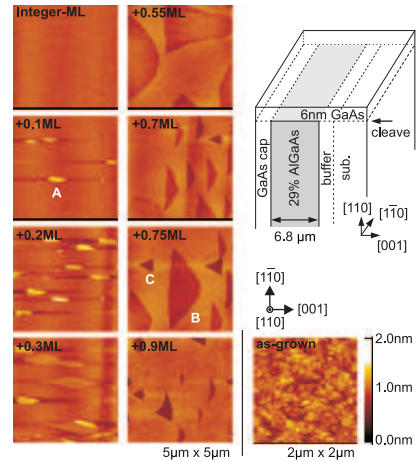}
\caption{
Atomic force microscopy (AFM) images of the surface of a 6-nm-thick (110) GaAs layer grown on a cleaved (110) edge of a 6.8-$\mu$m-thick Al$_{0.29}$Ga$_{0.71}$As layer on the (001) substrate. The observation area is 5 $\mu$m x 5 $\mu$m. A schematic sample structure is shown in the inset. The cleaved (110) surface is parallel to the major flat of the (001) substrate wafer. The (110) GaAs layer was overgrown at a substrate temperature of 490 ${}^{\circ}$C under As$_{4}$ flux. After the overgrowth, $\it{in}$ $\it{situ}$ annealing of the surface was done at a substrate temperature of 600 ${}^{\circ}$C for 10 minutes. A position with integral monolayer (ML) deposition is denoted as {\bf Integer-ML}, and other positions with fractional ML are denoted by deviations of deposition from the integral ML. As a reference, the AFM image of the surface without annealing is labeled as {\bf as-grown} with an observation area of 2 $\mu$m x 2 $\mu$m. 
}
\label{1}
\end{figure}

We first formed, by standard MBE on a (001) GaAs wafer, a 300 nm GaAs buffer layer, a 6.8-$\mu$m-thick Al$_{0.29}$Ga$_{0.71}$As barrier layer formed by repeating 20.09 nm Al$_{0.32}$Ga$_{0.68}$As layer and 2.55 nm GaAs healing layer with 15 s growth interruption, and a 3000 nm GaAs cap layer. On a cleaved (110) edge of this substrate, we performed a second MBE step growing a 6-nm-thick (110) GaAs layer by means of the cleaved-edge overgrowth method\cite{PfeifferAPL} (inset of Fig. 1). Though the overgrowth thickness was 6 nm on average, a spatial gradation of thickness of 1 \%/mm was intentionally introduced by aligning the surface along the Ga-flux gradient of MBE without substrate rotation to cause surface coverage variation by fractions of a monolayer for a surface atomic-step kinetics study.
The MBE growth conditions on (110) GaAs are unusual requiring a low substrate temperature of 490 ${}^{\circ}$C under a high As$_{4}$ overpressure.\cite{PfeifferAPL} After the overgrowth, the Ga flux was shuttered and an $\it{in}$ $\it{situ}$ anneal of the surface was performed at an elevated substrate temperature of 600 ${}^{\circ}$C for 10 minutes under the As$_{4}$ molecular flux.

Figure 1 shows surface morphology images by atomic force microscopy (AFM) on the GaAs annealed film together with a comparison image of an as-grown film without anneal. The as-grown surface is very rough, covered with many triangle-shaped island structures extending 50-100 nm laterally and 2-3 MLs in height. With the $\it{in}$ $\it{situ}$ annealing, however, the atomic-step and island densities are significantly reduced.\cite{note2} Note in the particular case where exactly 30 molecular MLs of GaAs was deposited over the cleave (denoted as {\bf Integer-ML}) that an atomically flat surface without any islands was formed over an area several tens of $\mu$m on a side. 

Notice also at other locations along the cleave where the GaAs coverage ended in a fractional monolayer, that atomic step-edges and islands or monolayer  pits are observed. When the deviation from an integer-ML thickness is +0.1ML or +0.2ML, the GaAs surface is generally atomically flat except for isolated islands shaped like $\it{boats}$ ({\bf A} in Fig. 1), which are 2- or 3-MLs high and elongated along the [001] direction. When the excess coverage becomes +0.3ML, larger islands of 1-ML height are found in addition to $\it{boats}$. At around +0.5ML coverage, the $\it{boats}$ disappear as the coalescence of 1-ML-height islands forms connected large terraces. For excess coverage of +0.7ML, a 1-ML-height terrace is extended over the whole surface with a few large isolated 1-ML-deep pits shaped like $\it{tropical}$ $\it{fish}$ facing toward the [001] direction ({\bf B} in Fig. 1). At still higher coverage the $\it{fish}$ shaped pits are joined by 2-MLs-deep pits shaped like $\it{arrowheads}$ pointing toward the [00\=1] direction ({\bf C} in Fig. 1). 
These characteristic shapes reveal the atomic-step kinetics during annealing and illustrate the strong driving force toward flat surface formation, which will be discussed elsewhere.

Note that the formed islands and pits are $\mu$m-scale in size, and thus are much larger than those observed on $\it{any}$ (001) GaAs MBE surface.\cite{HessSCI,GammonSCI,WuPRL} This suggests that during the anneal the Ga atom or GaAs molecular mobility is substantially higher on the (110) surface than on the (001) or other surfaces. Note also that these island and pit features are larger than the size of excitons in GaAs QWs and larger than the spatial resolution of an optical microscope.

\begin{figure}
\includegraphics[width=.30\textwidth]{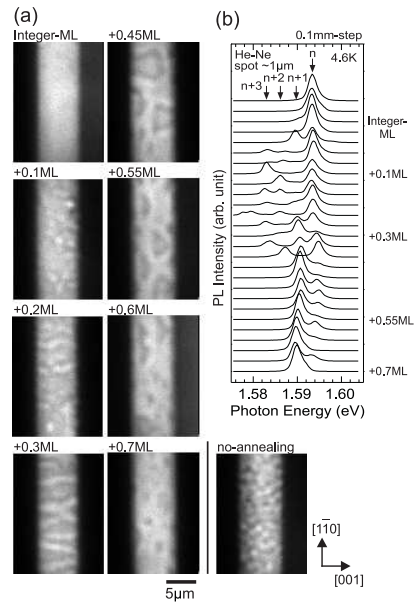}
\caption{
Microscopic photoluminescence (PL) imaging and spectroscopy obtained via the (110) surface in a backward scattering geometry.\cite{YoshitaJAP} (a) PL images of the GaAs QW at 4.7 K under uniform excitation of a He-Ne laser at well thickness positions up to +0.7ML. The spatial resolution of the images is 1 $\mu$m. For comparison, a PL image of the GaAs QW without annealing is shown as {\bf no-annealing}.  (b) PL spectra observed at the center of the 6.8-$\mu$m-wide QW region under point excitation of a He-Ne laser with a spot size of 1 $\mu$m. The excitation position was scanned with a step of 100 $\mu$m along the [1\=10] direction.
}
\label{2}
\end{figure}

To fabricate a QW with atomically flat interfaces, we overgrew an upper barrier on the top surface of Fig. 1 by continuing the cleaved-edge overgrowth. An Al$_{0.33}$Ga$_{0.67}$As barrier layer was grown at a substrate temperature of 490 ${}^{\circ}$C after the cleaved-edge overgrowth of the 6-nm-thick (110) GaAs well layer of 490 ${}^{\circ}$C and the $\it{in}$ $\it{situ}$ 10-minute anneal at 600 ${}^{\circ}$C. Figure 2 (a) shows micro-PL images of the GaAs QW at 4.7 K under uniform excitation in the same top-view geometry as in Fig. 1. The spatial resolution of the PL images is 1 $\mu$m.\cite{YoshitaJAP} Strong PL comes from the 6.8-$\mu$m-wide region on Al$_{0.29}$Ga$_{0.71}$As where the QW was formed. In the reference QW formed without annealing (denoted as {\bf no-annealing}), a PL image was spatially inhomogeneous, and had a spectral linewidth of 8 meV (full width at half maximum (FWHM)). This is consistent with a large roughness on the top interface, and previous studies on the related structures.\cite{HasenNAT,YoshitaPRB,GislasonAPL} On the other hand, in the QW with $\it{in}$ $\it{situ}$ 600 ${}^{\circ}$C annealing, a spectrally sharp (see below) and spatially uniform PL image was observed at the integer-30-MLs-thickness location (denoted as {\bf Integer-ML} in Fig. 2 (a)), which was also found to extend over several tens of $\mu$m in area. This demonstrates the formation of a 6-nm or 30-MLs GaAs QW with no monolayer steps or roughness at $\it{either}$ AlGaAs interface. 

Moreover, we see bright PL spots due to $\it{boats}$ at 30.2 MLs ({\bf +0.2ML} in Fig. 2 (a)), and dark PL profiles shaped like $\it{fish}$ at 30.55 MLs ({\bf +0.55ML} in Fig. 2 (a)). The AFM patterns of atomic steps observed in Fig. 1 are in fact reproduced in the PL image at each corresponding position of the fractional-MLs deposition. This confirms that the bottom interface formed by the cleavage has no atomic steps, and further that the surface morphology formed on the well layer during annealing and observed by AFM was conserved at the top interface of the QW as the upper barrier material was overgrown.

\begin{figure}
\includegraphics[width=.23\textwidth]{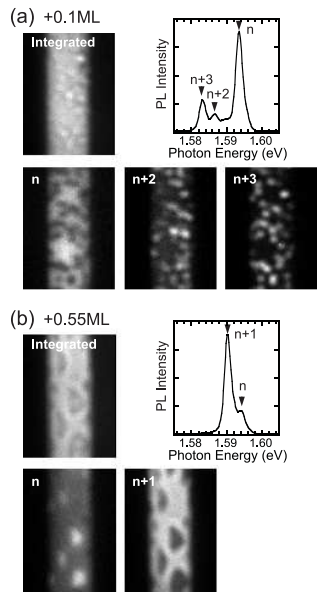}
\caption{
Integrated and spectrally resolved PL images of the QW and simultaneously obtained PL spectra at 4.7 K under uniform excitation at (a) +0.1ML and (b) +0.55ML thickness locations. The PL images are resolved at PL peak energies corresponding to the well thickness of n, n+1, n+2, and n+3 MLs.
}
\label{3}
\end{figure}

This conservation of the surface morphology at the QW hetero-interface and the resulting formation of the integral-MLs thick QW without interface roughness is also supported by spatially resolved PL spectroscopy associated with the spectrally resolved PL imaging. 
Figure 2 (b) shows PL spectra at various positions of integer-ML thickness and then as the thickness gradually increases up to an additional 0.7ML. At {\bf Integer-ML}, only a single PL peak (denoted as {\bf n}) forms as expected for a QW with integer-ML thickness. We observed a FWHM of 2 meV with spectral resolution of 0.8 meV.\cite{YoshitaUNP} This is much narrower than the FWHM of 8 meV observed for the reference QW without annealing, however we believe that even our annealed QW contains additional sources of broadening due to, for example, alloy scattering of the exciton wavefunction penetrating into the Al$_{x}$Ga$_{1-x}$As barriers where the Al alloy fraction is locally varying and/or interdiffusion of Al and Ga atoms happening at the top hetero-interface of the QW during the overgrowth of the AlGaAs barrier. We intend to examine these effects in a future report.

As well thickness is increased by +0.1ML to +0.3ML, emission peaks appear whose energy separation from the peak {\bf n} corresponds to a +2- or +3-MLs difference in well thickness, but a spectral peak from n+1 MLs does not. This is consistent with the formation of the $\it{boat}$-shaped islands of 2- or 3-MLs height that we observed in the AFM image. For GaAs depositions in excess of +0.3ML, on the other hand, the peaks {\bf n+2} or {\bf n+3} suddenly disappear and a peak corresponding to the n+1-MLs thickness appears in the PL spectra. This corresponds to the elimination of $\it{boat}$-shaped islands and formation of a large terrace of 1-ML height. 

Figure 3 shows spectrally resolved PL images at the (a) +0.1ML and (b) +0.55ML thickness positions, in which the intensities of the PL peaks denoted as n, n+1, n+2, and n+3 MLs are mapped. At the +0.1ML thickness position, the PL image is decomposed into bright spots due to $\it{boat}$-shaped islands at n+2- and n+3-MLs PL peaks and a reverse pattern from the surrounding n-MLs flat region. At the +0.55ML thickness position, the $\it{fish}$-shaped regions with n-MLs thickness are clearly resolved from the surrounding n+1-MLs region. 

It is interesting that the bright regions of {\bf n} are smaller in size than the corresponding dark regions of {\bf n+1} in Fig. 3 (b). This directly images the diffusion of quantum well excitons over the 0.8 $\mu$m distance from the higher-energy {\bf n} region image edge of the locally narrow QW of $\it{fish}$-shaped pits to the actual step edge of the larger lower-energy {\bf n+1} region. Such a long distance path at low temperature is not unreasonable as exciton diffusion is expected to be especially efficient in QWs with atomically smooth interfaces. 

The atomically smooth interface formation investigated here by means of the cleaved-edge overgrowth and $\it{in}$ $\it{situ}$ annealing method is expected to play an essential role in the realization and future investigation of high-quality quantum wells and wires with ideal 2-D and 1-D properties.
Indeed, by applying this annealing method, we have recently fabricated a high-quality T-shaped quantum wire, which shows a PL linewidth narrower by an order of magnitude than those previously reported and reveals striking 1-D physics.\cite{AkiyamaSSC}

In summary, by $\it{in}$ $\it{situ}$ growth-interrupt-anneal, we have controlled the surface morphology of the (110) GaAs epitaxial layer on the cleave and formed an atomically flat surface on layers of integral ML thickness. Using this technique, we have fabricated an exactly integral-MLs-thick GaAs QW in which neither barrier-well hetero-interface shows any atomic roughness, and which shows spatially uniform and sharp PL over an area extending several tens of $\mu$m.

The authors wish to thank Professor P. B. Littlewood in Cambridge University for his helpful discussions. Part of this research in Japan is supported by the Ministry of Education, Culture, Sports, Science and Technology, Japan.


\begin{references}

\bibitem{HessSCI}
H. F. Hess, E. Betzig, T. D. Harris, L. N. Pfeiffer, and K. W. West, Science {\bf 264}, 1740 (1994).

\bibitem{ZrennerPRL} A. Zrenner, L. V. Butov, M. Hagn, G. Abstreiter, G. B\"ohm, and G. Weimann, Phys. Rev. Lett. {\bf 72}, 3382 (1994).

\bibitem{BrunnerPRL}
K. Brunner, G. Abstreiter, G. B\"ohm, G. Tr\"ankle, and G. Weimann, Phys. Rev. Lett. {\bf 73}, 1138 (1994); Appl. Phys. Lett. {\bf 64}, 3320 (1994).

\bibitem{GammonSCI}
D. Gammon, E. S. Snow, B. V. Shanabrook, D. S. Katzer, and D. Park, Science {\bf 273}, 87 (1996); Phys. Rev. Lett. {\bf 76}, 3005 (1996).

\bibitem{WuPRL}
Q. Wu, R. D. Grober, D. Gammon, and D. S. Katzer, Phys. Rev. Lett. {\bf 83}, 2652 (1999).

\bibitem{PfeifferAPL} L. Pfeiffer, K. W. West, H. L. St\"ormer, J. P.
Eisenstein, K. W. Baldwin, D. Gershoni, and J. Spector, Appl. Phys. Lett.
{\bf 56}, 1697 (1990).

\bibitem{HasenNAT} J. Hasen, L. N. Pfeiffer, A. Pinczuk, S. He, K. W. West,
and B. S. Dennis, Nature {\bf 390}, 54 (1997).

\bibitem{YoshitaPRB} M. Yoshita, N. Kondo, H. Sakaki, M. Baba, and H.
Akiyama, Phys. Rev. B {\bf 63}, 075305 (2001).

\bibitem{YoshitaJJAP} M. Yoshita, H. Akiyama, L. N. Pfeiffer, and K. West, Jpn. J. Appl. Phys. {\bf 40}, L252 (2001).


\bibitem{note2} Gradual undulation of height along the horizontal direction in the background of all the images and dark stripes seen in front of and behind island structures in +0.1ML and +0.2ML are artifacts of the AFM background subtraction process, and should be ignored.

\bibitem{YoshitaJAP} M. Yoshita, H. Akiyama, T. Someya, and H. Sakaki, J. Appl. Phys. {\bf 83}, 3777 (1998).

\bibitem{GislasonAPL} H. Gislason, C. B. S$\o$rensen, and J. M. Hvam, Appl. Phys. Lett. {\bf 69}, 800 (1996).

\bibitem{YoshitaUNP} M. Yoshita, H. Akiyama, L. N. Pfeiffer, and K. W. West (unpublished).

\bibitem{AkiyamaSSC} H. Akiyama, L. N. Pfeiffer, A. Pinczuk, K. W. West, and M. Yoshita, Solid State Commun. {\bf 122}, 169 (2002).

\end{references}
\end{document}